\documentclass[preprint,preprintnumbers,prd,amsmath,amssymb,nofootinbib,tightenlines]{revtex4}
\usepackage{graphicx}
\usepackage{dcolumn}
\usepackage{bm}

\begin{document}
\preprint{FERMILAB-PUB-08-456-A-T}

\title{\boldmath%
Shedding Light on Dark Matter: A Faraday Rotation Experiment to Limit 
a Dark Magnetic Moment 
\unboldmath}

\author{Susan Gardner}

\affiliation{
Center for Particle Astrophysics 
 and Theoretical Physics, 
Fermi National Accelerator Laboratory, Batavia, IL 60510}

\affiliation{Department of Physics and Astronomy, University of Kentucky, 
Lexington, KY 40506-0055\footnote{Permanent Address} 
}


\begin{abstract}
A Faraday rotation experiment can set limits on the magnetic moment
of a electrically-neutral, dark-matter particle, and the limits increase in
stringency as the candidate-particle mass decreases. 
Consequently, if we assume the dark-matter particle to be a thermal relic, 
our most stringent constraints emerge 
at the keV mass scale. 
We discuss how such an experiment could be realized and determine 
the limits on the magnetic moment as a function of mass which 
follow given demonstrated experimental capacities. 
\end{abstract}

\maketitle

\section{Introduction}

Disparate astronomical observations provide compelling evidence for 
additional, non-luminous matter, or dark matter, in 
gravitational interactions.  
The evidence includes the persistence of the galactic rotation curves to 
distances for which 
little luminous matter is present~\cite{Faber:1979pp,rubin}, 
the relative 
strength and shape of the galaxy-distribution 
power spectrum at large wave numbers~\cite{galcluster}, and 
the pattern of acoustic oscillations in the power spectrum of the
cosmic microwave background~\cite{bao}. 
The cosmological evidence, in aggregrate, points to 
the assessment that dark matter comprises 
some twenty-three percent 
of the energy density of the universe,
with a precision of a few percent~\cite{concord}. 
Yet we know little about the nature of dark matter --- 
we do not know its mass,
its quantum numbers, or even with surety that it is indeed 
composed of isolated elementary particles. 
A recent gravitational lensing study does 
disfavor a modification of gravity in explanation of its 
effects~\cite{bullet}: 
we shall assume~\cite{Zhang:2007nk} that dark matter exists and is matter. 

We do know other things about dark matter~\cite{olivetasi,silk}; 
namely, that it is not hot~\cite{wfd}, 
and that it appears to lack both electric and color 
charge~\cite{ellis,echarge,echargeexp,echargeastro,ccharge,comment}. 
Here we use temperature, i.e., whether
dark matter is ``cold'' or ``hot,''
to connote whether dark matter is non-relativistic 
or relativistic, respectively, at the redshift
at which it decouples 
from matter in the cooling early Universe. 
For so-called thermal relics, this criterion selects 
the mass of the dark candidate as well,
so that colder particles are heavier. However, alternative production 
scenarios can exist, and very light particles
can also act as cold dark matter, as in the case of the axion~\cite{axioncdm}.
Nevertheless, model-dependent constraints do 
exist on the mass of the candidate particle, and we consider them 
in Sec.~\ref{masslimit}. 

The evidence for dark matter emerges from astrophysical 
observations of gravitational interactions, but 
establishing its couplings to Standard Model (SM) particles has
proven elusive. Nevertheless, such a quest is of great import, 
for it is through such means that 
its mass and quantum numbers can ultimately be determined. 
Indeed all direct and 
indirect means of detecting dark matter rely on the notion that
it does indeed experience weak or electromagnetic interactions 
to some degree. Turning to the known particles
of the SM for guidance, the neutron shows us that a particle can have 
both a vanishing electric charge and a 
significant magnetic moment. Thus we 
wish to constrain the possibility
that dark matter has a small electromagnetic coupling, 
via its magnetic moment. We review the existing
constraints on this possibility in Sec.~\ref{mulimit}. 

In this paper we consider a new technique for the direct
detection of dark matter, namely, through use of the 
gyromagnetic Faraday effect~\cite{svg}. Alternatively, 
this effect can be used to limit a possible 
magnetic moment $\mu$ of a dark-matter particle of mass $M$. 
Let us consider how this could work. 
An electrically neutral medium of particles which 
possesses a net magnetization in 
an external magnetic field is circularly birefringent, 
even if the medium is isotropic. 
This implies that the 
propagation speed of light in the medium depends 
on the state of its circular polarization, so that 
light prepared in a state
of linear polarization will suffer a rotation of
the plane of that polarization upon transmission through
the medium~\cite{polder,hogan}, 
as long as it does not travel at right angles
to the external magnetic field. We term this the 
gyromagnetic Faraday effect, after Ref.~\cite{EEbook}. 
Note that we need not rely on any existing magnetization
of the dark matter in order to realize an effect. 
Rather we imagine a 
Faraday rotation experiment mounted in a 
region with a large external 
magnetic induction $\bm{B}_0$ and dark matter 
of spin $S$ incident on it 
in the direction of $\bm{B}_0$. 
If the value of $\mu B_0$ 
is larger than the dark matter particle's kinetic energy in the
Earth's rest frame, then the field region 
acts as a spin-filter, or longitudinal Stern-Gerlach, 
device --- at least the highest energy 
spin configuration cannot enter the field region. 
This technique is used to polarize ultra-cold neutrons
(UCNs) in the UCNA experiment at Los Alamos
with near 100\% efficiency~\cite{Tipton:2000qi}. 
Thus, viewed in the Earth's rest frame, the dark matter 
which sweeps through the field region can possess a net magnetization. 
If light transits this medium in the direction of the magnetic field, 
and if we define $k_\pm$ to be the wave number 
for states with right- ($+$) or left-handed ($-$) circular 
polarization, then the rotation linearly polarized light
suffers in its transit through the medium is given 
by the angle $\phi = (k_+ - k_-)l/2$, 
where $l$ is the length of transmission through the medium. 
If $\phi$ is 
non-zero once all systematic effects which could mimic the
signal are excluded, 
we have evidence for dark matter with a non-zero 
magnetic moment. 

The direct detection of dark matter with a magnetic moment
could be realized in a variety of ways. 
For example, one could 
search for anomalous recoil events in scattering from nuclei, 
in just the manner one searches for spin-independent and spin-dependent 
dark matter-nucleon 
interactions~\cite{stodolsky,Smith:1988kw,Gaitskell:2004gd}. 
Indeed, an experimental signal of the latter, from the 
DAMA experiment~\cite{Bernabei:1996vj}, 
has been interpreted 
as a limit on a putative dark-matter magnetic moment, 
namely of
$\mu < 1.4\cdot 10^{-4} \mu_N$ 
for a dark matter candidate of 100 GeV in mass~\cite{Pospelov:2000bq}. 
In such experiments, however, the light dark-matter candidates we discuss 
give rise to momentum transfers which are much too small to be 
detected, even if the channeling effect proposed by Ref.~\cite{channel} 
and studied in Ref.~\cite{damastudy} 
is operative for the events studied in the 
DAMA/LIBRA NaI detector~\cite{Bernabei:2008yi} --- note 
Ref.~\cite{Petriello:2008jj} for further discussion. 
Alternatively, a dark matter magnetic moment can be found through either
nuclear magnetic resonance (NMR) or Faraday rotation studies. 
A NMR signal is typically realized through the detected change 
in magnetic field incurred through a spin-flip transition of the 
magnetic-moment-carrying particle, induced 
by an applied radio frequency. However, very small magnetic fields 
are more efficiently discovered through 
magnetoptical studies~\cite{SLR,BRoptmag}. Thus 
we focus on the use of Faraday rotation to detect dark matter. 
Precision optical rotation studies have also been conducted in which 
the external magnetic field is oriented at right angles to the direction of the
propagating light, to the end, e.g., of 
testing the optical birefringence of the vacuum, or,
alternatively, of limiting 
the photon-axion coupling~\cite{Cameron:1993mr,Zavattini:2005tm}.
Such an experimental configuration does not support a nonzero
Faraday effect, yet the empirical parameters 
used in these studies are 
useful to us, for we employ
them to estimate the limit on the magnetic moment as a function 
of mass in the set-up we consider. 

A great variety of dark matter candidates, consistent
with the various astrophysical constraints, exist in the
literature~\cite{Roszkowski:2004jc,dmsag}. Indeed, 
their masses vary from some $10^{-32}$ to $10^{15}$ GeV, 
and their interaction cross sections --- with nucleons --- vary over orders
of magnitude as well. Typical direct detection strategies
rely on the observation of ``anomalous'' 
nuclear recoils~\cite{dmsag,Smith:1988kw,Gaitskell:2004gd}, 
so that their sensitivity is typically to candidates of
${\cal O}$(100 GeV) in mass scale.
The experiment we suggest is sensitive to a completely different window
in parameter space --- to dark-matter candidates of
crudely ${\cal O}$(1 MeV) or less in mass. 
Although constraints can be set through
cosmological studies~\cite{svg}, terrestrial studies 
are amenable to better control, for the existence of cosmological magnetic
fields have not yet been established~\cite{Blasi:1999hu}.

Let us conclude our introduction by outlining 
the sections to follow. 
We begin by describing, in Secs.~\ref{masslimit}
and \ref{mulimit}, respectively, the 
mass and magnetic moment constraints which exist on 
a light dark-matter candidate with a non-zero magnetic moment.  
We then proceed to review the gyromagnetic Faraday effect
in Sec.~\ref{faraday} and 
to describe in concrete terms the experimental limits on $\mu$ one
might possibly attain in Sec.~\ref{setup}. We conclude with 
a summary in Sec.~\ref{summ}. 

\section{Mass Contraints}
\label{masslimit}

In standard Big-Bang cosmology, the nature of dark matter
impacts the formation of the large-scale structure of the Universe. 
In particular, if dark matter is cold and collisionless, then 
galaxy formation proceeds via a hierarchial clustering, namely, 
from the merging of small protogalactic clumps, on ever larger
scales~\cite{White:1977jf,fwd,pancake,Davis:1985rj}.
In contrast, if dark matter is hot, the hierarchy is inverted, so that 
large protogalactic disks form first, which then clump~\cite{fwd,pancake}. 
Galaxies, however, are observed at much larger redshifts than such simulations
predict~\cite{fwd,pancake}. Moreover, 
observations of particular classes of quasar absorption 
lines, the so-called damped Lyman-$\alpha$ systems,
thought to be the evolutionary progenitors of galaxies
today, also favor the former scenario~\cite{DLAS}. 
It has also been argued that hot dark matter, i.e., most notably, 
light, massive neutrinos, 
cannot explain the galactic rotation curves~\cite{tremaine}. 
However, the cold-dark-matter 
paradigm does have difficulties in confronting small-scale structure; 
it yields, in effect, 
too much clumpiness below the Mpc scale. 
Warm dark matter has been advocated as a way to 
alleviate these difficulties~\cite{wdm}. 
Limits on the mass of 
warm dark matter emerge from the comparison of the observations 
of the Lyman-$\alpha$ absorption spectrum with numerical 
simulations~\cite{viel1,Seljak:2006qw,Viel:2006kd,viel}; 
the limits depend on the particle considered, 
and the manner in which it is produced~\cite{petraki}, 
yielding~\cite{viel},  
at 2$\sigma$, $M \gtrsim 4$ keV for a thermal relic 
and $M \gtrsim 28$ keV for a massive sterile 
neutrino~\cite{Dodelson:1993je}. 

Cosmological constraints also exist on the mass of a 
dark-matter particle. If the particles annihilate via the 
weak interaction, then $\sigma_{ann} v$  
is parametrically set by ${\cal N}_AG_F^2 M^2$, 
where $G_F$ is the Fermi constant, 
${\cal N}_A$  is a dimensionless factor, 
and we assume 
$\sigma_{ann} \propto 1/v$.
In this case avoiding a dark-matter abundance in excess of 
the observed relic density bounds $M$ from below.  
Indeed, 
under these conditions the mass of the cold dark-matter particle must exceed 
${\cal O}(2\,\hbox{GeV})$ to avoid closing the Universe~\cite{mlimit}. 
The resulting  lower bound  on $M$ can be relaxed in different ways. 
Feng and Kumar~\cite{fengkumar}, e.g., have emphasized that 
the appearance of $G_F$ in $\sigma_{ann} v$ is simply parametric, 
that $G_F$ can be replaced with $g_{\rm eff}$, and that 
the effective coupling $g_{\rm eff}$ can be small without
having the precise numerical value of $G_F$. Thus if $g_{\rm eff} > G_F$, 
the bound on $M$ is weakened. 
Indeed, such considerations permit dark matter candidates which 
confront the relic density and big-bang nucleosynthesis 
constraints successfully but yet 
range from the keV to the TeV scale in mass~\cite{fengkumar,fengtuyu}. 

In this paper we consider what empirical constraints can be
placed on the magnetic moment of a dark matter particle. 
This hypothesis gives rise to a 
new annihilation mechanism, though both a dark matter particle and 
its antiparticle must be present to realize it. 
We recall that 
particles with magnetic moments
are invariably described by complex field representations, so that 
such a particle and its anti-particle are physically distinct ---
by the CPT theorem we expect the magnetic moments of such particles
to differ only in sign. 
If the particle-antiparticle annihilation is 
mediated by a magnetic moment interaction, 
then $\sigma_{ann} v$ 
is parametrically set by 
${\cal N}_A^\prime \alpha^2 {\mu}^2$ 
with ${\cal N}_A^\prime$ a dimensionless parameter,
as long as $M \gtrsim m_e$, the electron mass. 
The annihilation of still lighter
mass dark matter candidates follows a different parametric form. 
That is, if the particles 
annihilate to Standard Model particles, then $\sigma_{ann} v$ is suppressed
by higher powers of the coupling constant --- at least. Alternatively, 
if they annihilate to ``secluded'' dark matter particles, 
note, e.g., Ref.~\cite{kroli}, then 
$\sigma_{ann} v$ is of form ${\cal N}_A^\prime 
\alpha \alpha^\prime {\mu}^2$, where $\alpha^\prime$ is 
the electromagnetic coupling 
of the secluded particles and  ${\cal N}_A^\prime$ is a dimensionless parameter. 
Generally, we expect the magnetic moment $\mu$ to be of form 
$\mu=\kappa e \hbar/2M$, 
where $\kappa$ is the anomalous magnetic moment, 
so that $\sigma_{ann} v$ 
scales
as $1/M^2$. 
The presence of an additional annihilation mechanism should make the light 
dark matter candidates we consider, of ${\cal O}(\hbox{keV})$ scale 
in mass, say, 
decouple as matter and not radiation, so that 
constraints on the number of relativistic degrees of freedom during 
the epoch of big-bang nucleosynthesis also do not 
apply. This new annihilation mechanism becomes more effective 
as the candidate mass grows lighter. 
Nevertheless it is still possible to saturate 
the dark-matter density with such a candidate particle, 
for the efficacy of the annihilation process 
can be mitigated by a particle-antiparticle
excess~\cite{leptonasym,Sigurdson:2004zp}. 
Thus the dark matter relic density 
need not bound the magnetic moment from above. 
Moreover, 
we emphasize that dark matter could have multiple 
components, so that an upper bound on the magnetic moment could also
be evaded by diluting the magnetic-moment-carrying particle with
other sorts of dark matter.
We shall assume these various annihilation mechanisms are effective enough
to permit dark matter candidates as light as 
${\cal O}(1\,\hbox{keV})$ in mass. 

The intensity and morphology of galactic positron emission, as studied
by the INTEGRAL satellite~\cite{integral}, has prompted much discussion of 
dark matter candidates with electromagnetic 
interactions~\cite{boehm,Beacom:2004pe,Beacom:2005qv,Hooper:2007tu, 
Hooper:2008im,Bernabei:2008mv,fengkumar,Khalil:2008kp}, 
which are of ${\cal O}$(1 MeV) scale in mass, 
as well as of other possibilities, as, e.g., 
in  Refs.~\cite{Bertone:2004ek,Wang:2005cqa,Finkbeiner:2007kk,Pospelov:2007xh}. 
If the pattern of the INTEGRAL spectra are indeed explained
by dark matter, then additional constraints follow on its nature. 
For example~\cite{Beacom:2004pe,Beacom:2005qv}, 
observational constraints on the diffuse photon flux also impose limits on the
mass of the dark-matter candidate $\chi$ through internal bremsstrahlung 
corrections to the annihilation process $\chi \chi \to e^+ e^-$. 
Such constraints 
apply to our scenario as well; the upshot is that 
the dark particle's mass is limited 
to be less than a few MeV~\cite{Beacom:2004pe,Beacom:2005qv}. 
In our case internal bremsstrahlung
contributions can also be generated by the magnetic-moment-carrying particle. 
However, we note that soft photon emission via a M1 transition of 
an isolated magnetic dipole is slow compared to the rate set by the
inverse age of the universe, as we discuss in greater detail in 
Sec.~\ref{faraday}, so that 
no meaningful limit follows on its magnetic moment.
In what follows we consider candidate particles which range 
from ${\cal O}$(1 keV) to ${\cal O}$(100 MeV) in mass, though 
our limits are most effective at sub-MeV mass scales.

\section{Magnetic Moment Contraints}
\label{mulimit}

Various constraints on the magnetic moment of a dark-matter 
particle for masses in excess of $1$ MeV 
have been considered in Ref.~\cite{Sigurdson:2004zp}. 
We review these and more, in order to provide a context
for the direct detection experiment we suggest. 
In the mass window of interest to us, two experimental constraints are 
important --- one comes from precision electroweak 
measurements~\cite{Sigurdson:2004zp},  
and the other comes from 
low-energy $e^+ e^-$ collider data, 
namely from the process 
$e^+ e^- \to \nu\bar\nu \gamma$~\cite{Grotch:1988ac,Tanimoto:2000am}. 
This last constrains the magnetic moment of the invisible particle directly
and thus offers a constraint on a dark-matter magnetic moment as well. 
The authors of these studies use data at center-of-mass energies sufficient 
to produce a $\nu_\tau\bar\nu_\tau$ pair, 
given accelerator constraints on its mass~\cite{Grotch:1988ac} --- this 
easily includes our mass range of interest. 
Interpreting their results 
as a limit on the anomalous magnetic moment of the tau 
neutrino, the low-energy analyses 
conclude 
$\mu_{\nu_\tau} < 4 \cdot 10^{-6}\mu_B$ at 90\% CL~\cite{Grotch:1988ac}, 
and 
$\mu_{\nu_\tau} < 9.1 \cdot 10^{-6}\mu_B$ at 90\% CL~\cite{Tanimoto:2000am}
from distinct data sets. 
A more severe limit on the $\nu_\tau$ magnetic moment does exist~\cite{donut};
however, the nature of the $e^+e^-$ limits allows us to interpret them 
in a manner useful to our current study. 
For a discussion of how low-energy $e^+ e^-$ collider data can probe 
particular MeV dark matter models~\cite{boehm}, 
see Ref.~\cite{Borodatchenkova:2005ct}.

Precision electroweak measurements also constrain the
magnetic moment~\cite{Sigurdson:2004zp}. 
The quantity $\Delta \hat r$ 
captures radiative corrections 
to the relationship between the fine-structure constant 
$\alpha$, the Fermi constant $G_F$, and the $W^\pm$ 
and $Z$ masses, $M_W$ and $M_Z$~\cite{Marciano:1999ih}. 
The difference 
between the empirically determined value of $\Delta \hat r$
and that computed in the Standard Model 
provides a window $\Delta \hat r^{\rm new}$ to which a dark-matter 
particle can contribute. Following 
Refs.~\cite{Sigurdson:2004zp,Profumo:2006im}, we assume 
$\Delta \hat r^{\rm new}$ is given by 
the vacuum polarization 
correction to the photon self-energy from a dark-matter 
particle with a magnetic moment, with no other adjustments. 
We choose to study the quantity $\Delta \hat r$ as its uncertainty
is dominated by that in the running of $\alpha$~\cite{pdg2006}. 
Thus we consider~\cite{Marciano:1999ih} 
\begin{equation} 
M_W^2 = \frac{\pi\alpha}{\sqrt{2} G_F} 
\frac{1}{{\hat s}_z^2
(1 - \Delta\hat r)} \,,
\end{equation}
where ${\hat s}_z$ is computed in the $\overline{\hbox{MS}}$ scheme
and is $(1 - M_W^2/M_Z^2)$ up to small corrections. 
To compute $\Delta\hat r^{\rm new}$, we first 
recall the general form of the electromagnetic vertex
with Dirac and Pauli form factors~\cite{peskin}, namely, 
$\Gamma^{\mu}(k+q,k) = \gamma^\mu F_1(q^2) 
+ i \sigma^{\mu \nu} q_\nu F_2(q^2)/2M$,  where in our case 
$F_1=0$ and $F_2=\kappa$.  
Using the conventions of Ref.~\cite{peskin}, we 
introduce the polarization tensor
\begin{equation}
i \Pi^{\mu\nu}_{2,tt}(q) = 
\kappa^2 e^2 \int \frac{d^4 k}{(2\pi)^4} 
\hbox{tr} \left\{  
\sigma^{\mu \alpha} \frac{q_\alpha}{2M} 
\frac{(\slash{\!\!\!k} + m)}{(k^2 - M^2)} 
\sigma^{\nu \beta} \frac{q_\beta}{2M} 
\frac{(\slash{\!\!\!k} + \slash{\!\!\!q} 
+ m)}{((k + q)^2 - M^2)}  \,,
\right\} \,,
\end{equation}
with $\sigma^{\mu\nu}\equiv i[\gamma^\mu,\gamma^\nu]/2$. 
Noting
$ \Pi^{\mu\nu}_{2,tt}(q) = (q^2 g^{\mu\nu} - q^\mu q^\nu) \Pi_{2,tt}(q^2)$
and 
\begin{equation}
\Delta \hat r^{\rm new} = \Pi_{2,tt}(M_Z^2) - \Pi_{2,tt}(0) - M_Z^2 
\left(\frac{\partial \Pi_{2,tt}(k^2)}{\partial k^2}
\Bigg|_{k^2=0} \right)\,,
\end{equation}
we use standard techniques~\cite{peskin} to determine
\begin{equation}
\Delta \hat r^{\rm new} = 
- \frac{\kappa^2 \alpha}{4\pi} \int_0^1 dx 
\left\{ 
\left(1 + \frac{x(1-x) M_Z^2}{M^2}\right) \log 
\left( 1 - \frac{x(1-x) M_Z^2}{M^2}\right)
+ \frac{x(1-x) M_Z^2}{M^2}
\right\} \,
\label{rnew1}
\end{equation}
and, in the limit $a\equiv (M_Z/M)^2 \gg 1$, that~\cite{note} 
\begin{equation}
\Delta \hat r^{\rm new} \sim - \kappa^2 \frac{\alpha}{4\pi} 
\left( \frac{a}{6}\log a - \frac{a}{9} + O(1)\right) \,.
\label{rnew2}
\end{equation}
With $\Delta \hat r^{\rm new} < 0.0010$ at 95\% CL~\cite{pdg2006},  
we find, e.g., with $M=m_e$, the electron mass, that 
$|\kappa| < 4.1\cdot 10^{-6}$, whereas if 
$M=m_e/10$ that $|\kappa| < 3.4\cdot 10^{-6}$~\cite{svg}. 

A variety of astrophysical constraints exist on the magnetic moment of the
neutrino, and they can be adapted to our current case as well. 
They emerge, in particular, 
from the impact of the additional cooling
mechanism such would render on stellar evolution and 
lifetimes and on supernovae~\cite{Raffelt:1990yz,Heger:2008er}. 
An additional, albeit somewhat weaker, constraint comes 
from confronting element abundances with the predictions 
of big-bang nucleosynthesis,
to yield, e.g., $\mu_\nu < 2.9\cdot 10^{-10} \mu_B$~\cite{Dolgov:2002wy}. 
These constraints can be 
significantly weakened by the candidate particle's 
mass~\cite{Grotch:1988ac}; the ability to produce
particles of ${\cal O}(10 \,\hbox{keV})$ in mass and more 
in plasma at stellar temperatures is limited. 
In the case of big-bang nucleosynthesis, the constraints
on the magnetic moment of a massive tau neutrino are also 
weakened, though values as large as $\mu_\nu \sim 10^{-6} \mu_B$ 
are nevertheless excluded~\cite{Dolgov:2002wy}.

The constraints we have considered in this section
can be weakened by other means as well. 
Since they arise from the effects of particle production, 
the most economical mechanism 
is compositeness; to include this, we need only include a form
factor at each electromagnetic vertex. Thus we replace 
$\kappa \to \kappa/(1  - {M_Z^2}/{M_c^2})^{2}$, 
where $M_c$ is the compositeness scale,
in our earlier formulae.  Thus for 
$M=m_e/10$ our earlier bound of 
 $|\kappa| < 3.4\cdot 10^{-6}$~\cite{svg} from $\Delta \hat r$ 
relaxes to $|\kappa| < 1.5$ if $M_c = 2$ GeV. 
In this scenario, however, 
the electrically charged 
constituents may well give rise to other observable effects. 
One possibility which avoids this would be to give a known 
charged particle, such as an electron, a small
hidden sector interaction, 
so that it can help constitute dark matter, 
though its contribution to 
$\Delta \hat r$ has already been taken into account. 
We proceed to consider the manner in which direct constraints 
can be set on $\mu$. 

\section{Gyromagnetic Faraday Effect}
\label{faraday}

A medium of free electric charges in an external magnetic
field is circularly birefringent and gives rise to a Faraday
effect~\cite{jackson}, as long as the light
 doe not propagate
at right angles to the magnetic field. This effect has long been used
in radio astronomy to study the properties of the 
interstellar medium~\cite{radio}. 
A Faraday effect can also arise in an
electrically neutral medium in an external magnetic field, 
if the constituents carry magnetic
moments and if they are aligned by that magnetic field 
to give the medium some net magnetization~\cite{polder,hogan}. 
We consider the latter possibility exclusively. 

To derive the gyromagnetic Faraday effect,
we apply a magnetic induction $\bm{B}_0$ 
in a magnetizable medium 
with circularly polarized electromagnetic
waves propagating parallel to it. The external field 
induces a magnetization $\bm{{\cal M}}_{\rm tot}$, i.e., 
a net magnetic moment/volume, where 
$\bm{{\cal M}}_{\rm tot} = 
\bm{{\cal M}}_0 + \bm{{\cal M}}$ 
and $\bm{{\cal M}}_0$ results from $\bm{B}_0$ alone. 
The total magnetization of a medium at rest in the laboratory frame 
obeys the Larmor precession formula 
\begin{equation}
\frac{ d \bm{{\cal M}}_{\rm tot} }{dt} = 
\frac{g\mu_M}{\hbar}
\bm{{\cal M}}_{\rm tot} \times \bm{B}_{\rm tot} \,,
\label{larmor}
\end{equation}
so that $g\mu_M/\hbar$, noting $\mu_M\equiv e\hbar/2 M$ with $e>0$ for a
particle of mass $M$, is the gyromagnetic 
ratio of the magnetic-moment-carrying
particle. We note $\mu= S g \mu_M$, where $S$ is the spin of the particle. 
The gyromagnetic Faraday effect was first 
derived for a ferromagnetic material~\cite{polder,hogan}, 
for which use of the magnetic field $\bm{H}$ 
is appropriate. 
Since dark matter is only 
 weakly self-interacting at most, our hypothesized 
dark matter should be treated as a paramagnetic material --- so that we
employ the magnetic induction $\mathbf{B}$ throughout, though 
the use of $\bm{H}$ is also commonplace~\cite{schiff}. 
Corrections to the Larmor formula result if 
the medium's particles move at a significant fraction of
the speed of light, or if the particles possess 
a non-zero electric dipole moment~\cite{Bargmann:1959gz}. 
We shall neglect the latter possibility and, moreover, 
shall consider dark-matter candidates for which 
relativistic effects are ultimately small corrections. 

To determine the relativistic corrections to Eq.~(\ref{larmor}), 
we first construct the covariant classical equation of motion
for a single spin in homogeneous electromagnetic fields. 
This is germane as we can and indeed do 
neglect the mutual interactions of the dark-matter particles, so that 
the magnetization is given by the quantum-mechanical
expectation value of the spin operator for a single particle times
the number density; and the time evolution of 
the expectation value is itself described by that of the 
associated classical equation of motion. The latter, for a 
charged particle with a spin, is given by the 
Bargmann-Michel-Telegdi equation~\cite{Bargmann:1959gz}.
We cannot use this result directly because no forces act on 
electrically neutral particles 
in homogeneous electromagnetic fields, so that no 
Thomas precession term is present~\cite{rauchwerner}. 
Nevertheless, well-known treatments~\cite{jackson} can be readily adapted
to this case. 
Requiring $d{\cal U}^\alpha/d\tau = 0$, where 
${\cal U}=c\gamma(1,\bm{\beta})$ is the 4-velocity of the particle 
in the laboratory frame 
and ${\cal S}$ is its spin, namely 
${\cal S}=({\cal S}^0,\bm{{\cal S}})$, 
we find that
\begin{equation} 
\frac{d{\cal S}^\alpha}{d\tau} 
= \frac{g \mu_M}{\hbar} \left( 
F^{\alpha\beta} {\cal S}_\beta + 
\frac{U^\alpha}{c^2} \left({\cal S}_\lambda F^{\lambda \mu} U_{\mu}\right)
\right)\,,
\end{equation}
where $F^{\alpha \beta}$ is the field-strength tensor in 
SI units and $\tau$ is the proper time of the particle. 
Thus in the laboratory frame the magnetization evolves as  
\begin{equation}
\gamma 
\frac{d \bm{{\cal M}}_{\rm tot} }{dt} = 
\frac{g\mu_M}{\hbar} 
\left(
\bm{{\cal M}}_{\rm tot} 
\times ({\bm{B}}_{\rm tot} - 
\bm{\beta} \times \frac{{\bm{E}}_{\rm tot}}{c})
+ \gamma^2 \bm{\beta} 
(\bm{\beta} \times \bm{{\cal M}}_{\rm tot} )\cdot ({\bm{B}}_{\rm tot} - 
\bm{\beta} \times \frac{{\bm{E}}_{\rm tot}}{c})
\right)\,,
\label{larmorbeta}
\end{equation} 
where we emphasize 
$\gamma$ is the Lorentz factor, 
namely $\gamma\equiv 1/\sqrt{1-\beta^2}$. To proceed, 
we separate $\bm{B}_{\rm tot}$ and $\bm{E}_{\rm tot}$ 
as $\bm{B}_{\rm tot}=\bm{B}_0 + \bm{B}$ 
and $\bm{E}_{\rm tot}=\bm{E}_0 + \bm{E}$, so that 
$\bm{{\cal M}}_0$ results exclusively from the external electromagnetic fields
$\bm{B}_0$ and $\bm{E}_0$. Working to leading order in the small quantities 
$\bm{{\cal M}}$, $\bm{B}$, and $\bm{E}$, which arise in the 
presence of electromagnetic radiation, 
we have 
\begin{eqnarray}
\gamma 
\frac{d \bm{{\cal M}} }{dt} &=&
\frac{g\mu_M}{\hbar} 
\Bigg(
\bm{{\cal M}}_0 
\times ({\bm{B}}_{0} - 
\bm{\beta} \times \frac{{\bm{E}}_{0}}{c}) + 
\bm{{\cal M}} 
\times ({\bm{B}}_{0} - 
\bm{\beta} \times \frac{{\bm{E}}_{0}}{c}) + 
\bm{{\cal M}}_0 
\times ({\bm{B}} - 
\bm{\beta} \times \frac{{\bm{E}}}{c}) \nonumber\\
&&+ \gamma^2 \bm{\beta} 
\Big(
(\bm{\beta} \times \bm{{\cal M}}_0)\cdot ({\bm{B}}_{\rm tot} - 
\bm{\beta} \times \frac{{\bm{E}}_{\rm tot}}{c})
+ 
(\bm{\beta} \times \bm{{\cal M}})\cdot ({\bm{B}}_{0} - 
\bm{\beta} \times \frac{{\bm{E}}_{0}}{c}) 
\Big)
\Bigg)\,.
\label{larmorbeta2}
\end{eqnarray} 
We can consider the evolution of the dark matter magnetization
in vacuum, i.e., in the absence of ordinary matter, or in matter. 
Since the largest external fields we can apply obey
${B}_0 \gg {E}_0/c$ in vacuum, we set $E_0=0$ henceforth. 
We note, however, that the atomic-scale separation of electric charges
in matter permit the opposite limit, 
${B}_0 \ll {E}_0/c$, so that the analysis of the magnetization 
in that case can be altogether distinct. We set this possibility
aside for later discussion and continue with the analysis in vacuum. 
We note that the ability to establish a vacuum relies on the
presence of matter with 
conventional electromagnetic and strong couplings; dark matter
is sufficiently weakly interacting that vacuum technology does not affect it. 
Hence we use ``vacuum'' to connote the absence of ordinary
matter. In this case we apply $\bm{B}_0$ 
in the $\hat{\bm{x}}$-direction and choose
$\bm{\beta}$ to be parallel or antiparallel 
to $\hat{\bm{x}}$ as well. As we have mentioned, 
the entry of dark matter into the 
magnetic field region acts as a spin filter
device. The dark matter which does enter the 
apparatus can thus possess a net magnetization, so that $\bm{{\cal M}}_0$ is in
the $\hat{\bm{x}}$-direction\footnote{This follows irrespective of
the sign of $g$. For $g{\stackrel{>}{<}}0$, however, the spins preferentially 
point in the $\pm\hat{\bm{x}}$ direction.}. As a result Eq.~(\ref{larmorbeta2}) reduces
to 
\begin{equation} 
\gamma 
\frac{d \bm{{\cal M}} }{dt} =
\frac{g\mu_M}{\hbar} 
\left(
\bm{{\cal M}} 
\times {\bm{B}}_{0} 
+ 
\bm{{\cal M}}_0 
\times ({\bm{B}} - 
\bm{\beta} \times \frac{{\bm{E}}}{c}) 
\right)\,.
\label{larmorvac}
\end{equation} 
Choosing the wave vector $\bm{k}$ of the
light in the $\hat{\bm{x}}$-direction as well, we recall 
$\bm{E} = -c \hat{\bm{x}}\times \bm{B}$ and let 
$\bm{B}(\bm{x},t) = B_\pm \bm{e}_{\pm} 
\exp(i k_\pm x -i \omega t)$, 
where $\bm{e}_\pm \equiv \hat{\bm{y}} \pm i \hat{\bm{z}}$. 
We define the polarization state with positive helicity, $\mathbf{e}_+$, 
to be right-handed, which differs from the convention used in optics. 
In steady state, we find $\bm{{\cal M}}= {\cal M}_\pm \bm{e}_{\pm}$ and finally
that 
\begin{equation}
{\cal M}_\pm = \pm  \frac{\omega_M (1+\beta)}{\gamma\omega \pm \omega_B} B_\pm 
\equiv \chi_\pm B_\pm \,,
\label{solnvac}
\end{equation}
where we have chosen $\bm{\beta}=-\beta \hat{\bm{x}}$ and defined
$\omega_M\equiv g\mu_M {\cal M}_0/\hbar$ and $\omega_B\equiv g\mu_M B_0/\hbar$. 
Since $k_\pm =(\omega/c)\sqrt{1 + \chi_\pm}$, we have 
\begin{equation}
k_\pm = \frac{\omega}{c} \sqrt{1 \pm 
\frac{\omega_M(1+\beta)}{\gamma \omega \pm \omega_B}} \,,
\label{kpm}
\end{equation}
or that 
\begin{equation}
k_+ - k_- = \frac{\omega_M (1+\beta)}{\gamma c}
\left( 1 + {\cal O}\left(
\frac{\omega_B^2}{\gamma^2\omega^2}\right)\right)\,,
\label{kdiff}
\end{equation}
where we note for conceivable light sources that 
$\omega \gg \omega_B, \omega_M$. 
Physically, the magnetic field associated with the passing
electromagnetic wave tugs on the spinning particle in a direction
perpendicular to ${\cal M}_0$, prompting it to emit radiation
which interferes with the light traveling in the forward direction,
generating the birefringence. 
The Faraday rotation angle $\phi$ is simply 
$\phi = (k_+ - k_-)l/2$, where $l$ is the total distance
travelled by the photon. The quantity $\omega_M$ is signed, so that 
the sense of the rotation angle determines the sign of $g$. 
If $\beta=0$, we recover the result of Ref.~\cite{svg}, 
whereas in the extreme relativistic limit, i.e., as 
$\gamma\to\infty$, $\chi_\pm \to 0$ and thus $\phi$ 
vanishes as well. 
The average value of $k_\pm$ is not altered to 
leading order in small quantities, namely, 
\begin{equation}
k_{\rm avg} \equiv  \frac{1}{2}(k_+ + k_-)  
= \frac{\omega}{c} 
\left(1 
+ {\cal O}\left(\frac{\omega_B \omega_M}{\gamma^2\omega^2}\right) 
\right) \,, 
\end{equation}
so that an appreciable Faraday rotation can accrue in 
the absence of an effect on the average group velocity. 
Thus far we have considered the Faraday rotation of linearly
polarized light consequent
to passage a distance $l$ through a medium; practical
considerations demand that we 
determine its properties under reflection as well. If
we reverse the direction of the light, Eq.~(\ref{solnvac})
is unaltered save for the sign of the term in $\beta$. The last does
change sign since $\bm{E} = -c \hat{\bm{k}}\times \bm{B}$. 
Thus if we set $\beta=0$, an initially right-handed circularly
polarized wave, e.g., travels both 
forward {\em and} backward 
with wave number $k_+$, so that the rotation angle accrues coherently
under momentum reversal. The additional Faraday rotation associated
with the explicit $\beta$-dependent term in Eq.~(\ref{kpm}), however, 
cancels under a round-trip transit. Such contrasting 
behavior is long familiar from 
the study of birefringence in 
chiral media~\cite{EEbook}, which break macroscopic parity invariance. 
Similar conclusions have been drawn from an analysis of 
parity-violating photon--external-field interactions
as well~\cite{Hu:2007vx}. 
Thus the net rotation angle 
after a round-trip, or after many, of a {\em total} travel length $l$ is 
\begin{equation}
\phi_0 = \frac{\omega_M l}{2\gamma c}
\left( 1 + {\cal O}\left(
\frac{\omega_B^2}{\gamma^2\omega^2}\right)\right)\,. 
\label{phi0}
\end{equation}
Neglecting the ${\cal O}(\omega^{-2})$ corrections and working
in the $\beta\to 0$ limit, this becomes simply 
\begin{equation}
\phi_0 = \frac{g \mu_M {\cal M}_0 l}{2\hbar c}\left(1 + {\cal O}\left(\beta^2
\right)\right)
\,,
\label{phisma}
\end{equation}
which agrees with the result of the non-relativistic 
treatment in Ref.~\cite{svg}. 
For a system at rest in thermal equilibrium,
the magnetization ${\cal M}_0$ is a simple function of 
the applied magnetic field. We recall that 
for a system of spin $1/2$ particles, e.g., each with magnetic
moment $\mu$, the magnetization for a system with number density 
$n_M$ at temperature $T$ is~\cite{polreview} 
\begin{equation}
{\cal M}_0 = n_M \mu \tanh \left( 
\frac{\mu B_0}{k_B T} 
\right) \,, 
\label{magnet}
\end{equation} 
though it is of little practical relevance to the current circumstance,
for the system we consider approaches 
thermal equilibrium extremely slowly. That is, 
a dilute gas in an external magnetic field polarizes through spontaneous
emission, and the rate $W$ for this process is given by that of a magnetic
dipole transition~\cite{sobelman}: 
\begin{equation}
W = \frac{4}{3\hbar} \left(\frac{\omega}{c}\right)^3 \left(g\mu_M\right)^2 \,,
\end{equation}
where $\omega = g \mu_M B/\hbar$. Thus even if 
$\mu_M = \mu_B$, the Bohr magneton, and $B=25$ T, we would have 
$W \sim 1\cdot 10^{-6}\,\hbox{s}^{-1}$, which is trivial compared
to the average rate with which dark matter is expected
to transit an experimental apparatus. 
Although dark matter
may possess some primordial magnetization, it is
likely so small~\cite{svg} that it is important to realize other means
of polarizing it. For the particular geometry we consider, as
we have noted, 
the onset of the 
magnetic field region acts as a spin filter
device. Although this method should yield some net magnetization for 
any non-zero spin $S$, we explicitly assume a 
spin $1/2$ candidate in what follows. 
If the velocity $\bm{v}_M$ 
of the incoming particles is aligned with the direction of 
the magnetic field, then particles
with $v_M \equiv \beta c  < v_{\rm stop}$ can only enter the magnetic field
region if their magnetic moment is aligned with it, 
where $v_{\rm stop}$ is such that 
\begin{equation}
\frac{1}{2} M v_{\rm stop}^2 = |\mu| B_0 \,. 
\label{vstop}
\end{equation} 
We have set any external electric field to zero
and have neglected corrections of ${\cal O}(\beta^2)$. 
The dark matter which does enter the 
apparatus can thus possess a net magnetization; namely, 
\begin{equation}
{\cal M}_0 = n_M \mu {\cal P}\,,
\label{Mcalc}
\end{equation}
where  ${\cal P}$ 
is the polarization of the spins. 
We define ${\cal P}\equiv (N_+ - N_-)/(N_+ + N_-)$,
where $N_+$ and $N_-$ are the number of spins pointing
in and against the direction of $\bm{B}_0$, respectively. If $\mu \to -\mu$
then ${\cal M}_0\to {\cal M}_0$ just as in Eq.~(\ref{magnet}). 
We study
the value of ${\cal P}$ as a function $\mu B_0$, $M$, and astrophysical 
parameters in the next section. 

Before proceeding, we return to the notion of studying
the Faraday rotation of dark matter passing through ordinary matter. 
In this regard, we wish to consider matter comprised of atoms with 
closed electron shells, 
so that there are no 
unpaired electrons present to engender a gyromagnetic Faraday effect. 
The exceptionally large electric fields associated with atoms and nuclei~\cite{fermi}
\footnote{In H-atom, e.g., 
a test charge a Bohr radius away from the proton sees 
$E/c \approx 2\cdot 10^3$ T.}, 
make it possible for $|\bm{\beta}\times \bm{E}_0|/c$ to exceed 
presently achievable external magnetic fields~\cite{record}. 
Such considerations yield 
significant limits on the neutron electric dipole moment~\cite{purcell}, 
e.g., from neutron-noble gas scattering~\cite{fermi,rabi}. 
Returning to Eq.~(\ref{larmorbeta2}), we choose 
$\hat{\bm{x}} \parallel \hat{\bm{k}}$ as in previous case, but now 
choose 
$\bm{\beta} \perp \hat{\bm{x}}$ so that $\bm{\beta}\times \bm{E}_0$ can
also be in the $\hat{\bm{x}}$ direction. 
Counting $|\bm{\beta}\times \bm{E}_0|/c$ 
as a parameter of ${\cal O}(1)$ and neglecting terms of 
${\cal O}(\beta^2)$ and
higher, we have 
\begin{equation}
\frac{d \bm{{\cal M}} }{dt} =
\frac{g\mu_M}{\hbar} 
\left(
-\bm{{\cal M}} 
\times \left( {\bm{\beta}}\times \frac{\bm{E}_0}{c} \right)
+ 
\bm{{\cal M}}_0 
\times \left({\bm{B}} - 
\bm{\beta} \times \frac{{\bm{E}}}{c}\right) 
\right)\,,
\label{larmormed}
\end{equation} 
and the steady-state solution 
\begin{equation}
{\cal M}_\pm = \pm  \frac{\omega_M }{\omega \mp \omega_E} B_\pm \,,
\label{solnmed}
\end{equation}
where $\omega_E\equiv g\mu_M |\bm{\beta}\times \bm{E}_0|/\hbar c$, 
which yields the rotation angle
\begin{equation}
\phi_0 = 
\frac{g \mu_M {\cal M}_0 l}{2\hbar c}
\left( 1 + {\cal O}\left(
\frac{\omega_E^2}{\omega^2}\,, \beta^2\right) \right)\,, 
\label{phi0mat}
\end{equation}
irrespective of whether round-trip paths are executed by the light. 
In this context, then, the effective magnetic field is relevant
simply to the value of ${\cal{M}}_0$. 
Here, too, we need
to determine the polarization of the dark matter which penetrates
the material. Suppose $\bm{\beta}=\beta \hat{\bm{y}}$ and that
it is possible to choose a material for which 
${E}_{0\,z} \gg E_{0\,x}$, so that 
the effective magnetic field is in the $\hat{\bm{x}}$ direction. 
The force
on the dark-matter particle in entering the medium is 
$\bm{\nabla} (\bm{\mu}\cdot (\bm{\beta}\times \bm{E}_0/c))$;
since we can expect the magnitude of $\bm{E}_0$ to depend on $y$,
a longitudinal Stern-Gerlach effect is still possible in 
this geometry. A force in the $z$ direction can engender
the more familiar transverse Stern-Gerlach effect, but 
the increasing diameter of the laser beam as a result
of scattering in its passage through the material 
may make it impossible to exploit this feature. 
Thus we tentatively conclude that it ought be 
possible to realize a meaningful Faraday rotation study 
in matter as well, with potential gains in sensitivity
to a possible dark matter magnetic moment. 
We now return to the vacuum case, to describe how 
such an experiment can be realized and to estimate the limits
on $\mu$ one could possibly obtain. 

\section{A Faraday Rotation Experiment}
\label{setup}

Although  
the Faraday rotation effect we discuss can be found through correlation 
studies of the polarization of the cosmic microwave 
background~\cite{svg}, a terrestrial Faraday rotation experiment 
offers a number of advantages. Current bounds on primordial
magnetic fields~\cite{Blasi:1999hu} make any primordial magnetization associated
with dark matter small~\cite{svg} and 
difficult to probe, 
particularly in experiments executed over terrestrial length scales. 
However, in this case, as we shall demonstrate, 
one can apply a very strong magnetic field of known strength, 
and polarize the dark matter to an appreciable degree. 
Moreover, as we have seen, the 
Faraday rotation associated with a magnetic moment 
accrues coherently under momentum reversal, 
so that the measurement can be made in a small cavity and
yet have a long effective path length. 
Measurements of very small rotation angles are also possible; 
the recent PVLAS experiment, for example, was able to achieve sensitivity  
to rotation angles of ${\cal O}(10^{-8})$~\cite{Zavattini:2005tm} --- this 
stands in constrast to a sensitivity of 
${\cal O}(10^{-4})$ anticipated with the future CMBpol 
satellite~\cite{planck}.

\begin{figure}
\includegraphics[scale=0.5,angle=0]{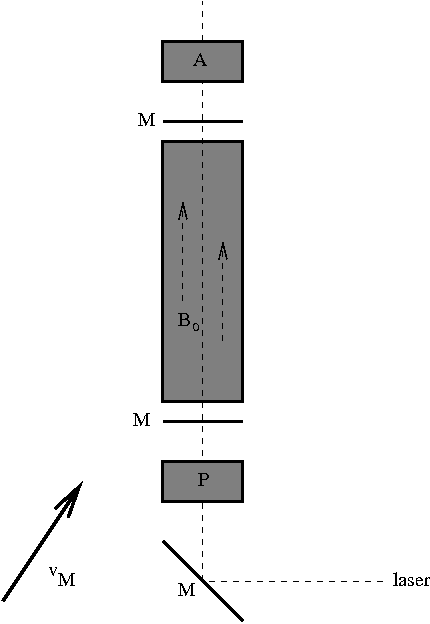}
\caption{\label{fig:schem} Schematic of the Faraday 
rotation experiment described in text.}
\end{figure}

A schematic 
of the Faraday rotation experiment we propose 
is illustrated in Fig.~\ref{fig:schem}. 
Its ingredients comprise a laser, 
a linear polarizer, 
an evacuated optical cavity through which the light makes 
multiple passes and to which 
a longitudinal, steady magnetic field has been applied, and an 
analyzer. The technical requirements of a sensitive Faraday rotation
measurement may demand a more sophisticated setup; however,  
such details are not needed for our estimate of the limit on 
$\mu$ for a given sensitivity to the rotation angle. 
We emphasize that since dark matter carries neither 
electric charge nor suffers strong interactions~\cite{silk}, 
it is unaffected by vacuum pumps and, indeed, can pass
into cavities free from ordinary matter. 
The generic setup we propose, 
save for the nature of the 
magnetic field, is common to the PVLAS 
experiment~\cite{Zavattini:2005tm}, which investigated
the optical properties of the vacuum,  as well. 
As that experiment was sensitive to extremely small changes in 
the photon polarization, 
we adopt it as a reference --- 
we use certain of the parameters chosen in that experiment in order to 
estimate the achievable bounds on the magnetic moment 
as a function of the mass of the 
dark-matter candidate. 
A remark concerning the orientation of the
apparatus in our schematic is in order. 
In our derivation of Eq.~(\ref{phisma}) we chose 
$\hat{\bm{\beta}}=\hat{\bm{x}}$, though our result is of more 
general validity. That is, if we return to Eq.~(\ref{larmorbeta2}),
setting $\bm{E}_0=0$ and working to ${\cal O}(\beta^2)$, 
we see that the absence of the 
${\cal O}(\beta)$ term in Eq.~(\ref{phisma}) 
results from the transformation properties 
of $\bm{E}$ under momentum reversal --- regardless
of the direction of $\bm{\beta}$. Consequently, we need not orient
the direction of the magnetic field with respect to the 
dark matter ``wind'' in any particular way, as we indicate
in Fig.~\ref{fig:schem}. 
It will turn out, however, that
the choice of the orientation of the apparatus 
with respect to the Earth's velocity 
can modify the efficacy of 
the longitudinal Stern-Gerlach device.

In order to evaluate the 
sensitivity of the scheme we suggest to a
dark matter magnetic moment, we need to make assumptions
concerning its local velocity distribution and number density. 
Both of the latter quantities are relevant to the computation
of ${\cal M}_0$ via Eq.~(\ref{Mcalc}), which gives rise 
to the rotation angle in Eq.~(\ref{phisma}). 
To do this we adopt 
the ``canonical model''~\cite{golwala,Lewin:1995rx}
employed in the analysis of direct detection 
experiments~\cite{stodolsky,Smith:1988kw,Gaitskell:2004gd} 
for cold dark matter. 
That is, we assume that the dark matter in our galaxy resides in 
a non-rotating halo and that the velocity distribution function 
$f(\bm{v})$ in that halo is that of an 
isothermal sphere. This assumption is motivated by simplicity.
More realistic 
distributions can affect the expected event rates, as well as 
their temporal variation~\cite{Green:2002ht,Green:2003yh,Fornengo:2003fm}. 
We note that $\bm{v}=\bm{v}_M + \bm{v}_E$, where $\bm{v}_M$
is the velocity of dark matter relative to the Earth, which 
we introduced earlier,  
and $\bm{v}_E$ is the velocity of the Earth relative 
to the nonrotating halo of the galaxy. 
The value of $v$ 
can range up to the galactic escape velocity, roughly 
$650\,\hbox{km/s}$~\cite{golwala},
though we do not impose this cut-off in what follows, 
because the normalization of the resulting distribution
differs by less than 1\%~\cite{Lewin:1995rx}. 
Thus the form of $f$ is that of a 
 Maxwell-Boltzmann 
distribution: 
\begin{equation}
f(\bm{v}_M,\bm{v}_E) = 
\frac{1}{\pi^{3/2} v_0^3} 
\exp\left(-\frac{(\bm{v}_M + \bm{v}_E)^2}{v_0^2}\right)\,. 
\label{maxwell}
\end{equation}
The velocity $v_0$ is related to the root-mean-square velocity of the
distribution; for a galaxy with a flat rotation curve 
it is argued to be equal to the radial velocity of the galactic 
disk~\cite{Drukier:1986tm} and thus 
in practice is taken to be $v_0\approx 220\,\hbox{km/s}$~\cite{golwala}. 
The velocity $\bm{v}_E$ is determined by the sum of 
\begin{equation}
\bm{v}_E= \bm{u}_r + \bm{u}_{s} + \bm{u}_{e} \,,
\end{equation}
where 
$\bm{u}_r$ is the velocity of the galactic disk in 
an inertial reference frame, 
$\bm{u}_{s}$ is the velocity of the Sun with respect to the galactic disk, 
and $\bm{u}_{e}$ is the velocity of the Earth about the Sun. 
Assuming the Milky Way is axisymmetric, 
$\bm{u}_r$ is fixed by the circular velocity at the Sun's radius
from the galactic center, which is 
$220\pm 20$ km/s~\cite{Kerr:1986hz,pdg2008}. 
In galactic coordinates 
$(-\hat{\bm{r}},\hat{\bm{l}},\hat{\bm{z}})$~\cite{Dehnen:1997cq}, 
in km/s, 
we thus employ~\cite{golwala}: 
$\bm{u}_r = (0,220,0)$, which follows if the
Milky Way is axisymmetric, and $\bm{u}_s = (9,12,7)$. The 
approximate speed of the Earth about the Sun is 30 km/s, whereas 
its approximate speed about its axis is 0.5 km/s. We ignore
the effects of the Earth's rotation about its axis in our analysis, 
as it is no larger than the error in $\bm{u}_s$~\cite{Dehnen:1997cq,binney}. 
Moreover, $\bm{v}_E$ is dominated by motion along the longitudinal
coordinate $\hat{\bm{l}}$. Noting this and that the ecliptic is oriented
at an angle of roughly $60^\circ$ with respect to the galactic 
equatorial plane 
means we can approximate 
the piece of $\bm{u}_E$ in the longitudinal direction $\hat{\bm{l}}$ 
as 
$30 \cos 60^\circ \cos (2\pi((t - 152.5)/365.25))$, 
to estimate 
$\bm{v}_E\approx (232 + 15 \cos (2\pi((t - 152.5)/365.25)))
\hat{\bm{l}}$~\cite{golwala}. We retain this approximation 
for simplicity, though the presence of the other components
of $\bm{v}_E$, as well as a 
more realistic velocity distribution, are important to an 
assessment of the event rates, the size and phase of
any temporal 
variations therein, 
and dark-matter 
exclusion limits to 
better than 
${\cal O}(100\%)$~\cite{Green:2002ht,Copi:2002hm,Green:2003yh}. 
We note, too, that more precise determinations of the 
astronomical inputs also exist~\cite{Dehnen:1997cq,binney}.

Finally, to complete our description of the model, we 
choose a dark matter mass density of 
$\rho = 0.3\, \hbox{GeV/cm}^3$~\cite{golwala}. 
The currently accepted range for $\rho$ is 
$0.2 - 0.4\, \hbox{GeV/cm}^3$~\cite{binneytr,Jungman:1995df} 
 for a smooth matter distribution with a 
spherical halo. Models of galaxy formation 
which relax the smoothness
assumption can give rise to local dark matter densities 
both larger and smaller than this range~\cite{Kamionkowski:2008vw}, 
where we refer to Ref.~\cite{Ellis:2008hf} 
for a succinct summary of the possibilities. 
Some of the uncertainties in the model we outline
can be correlated. 
For example, an elliptical halo, and concomitant triaxial velocity
distribution, can yield somewhat higher local 
densities~\cite{Gates:1995dw,Kamionkowski:1997xg}. Unfortunately, 
direct information on the dark-matter mass density in our solar system
is sparse, and observational bounds exceed the 
estimate we employ 
by orders of magnitude~\cite{Khriplovich:2006sq, Khriplovich:2007qt, 
Iorio:2006cn, Sereno:2006mw, Frere:2007pi, Xu:2008ep}. 

To realize a limit on $\mu$, we 
assert in what follows that dark matter is comprised of 
a single type of spin $1/2$ particle with fixed $M$ and $\mu$,
though we emphasize that the detection of a signal does
not require that the particle have spin $1/2$. 
Our numerical estimate is in two distinct parts. 
Operating in the canonical model, we 
first assess the polarization 
of the dark matter in the magnetic field region
as function of $M$ and $\mu$. With this in hand, 
we can then determine the limit on $|\mu|$
with $M$
which follows from a given limit on the magnitude of
the Faraday rotation angle. 

\begin{figure}
\includegraphics[scale=0.5,angle=-90]{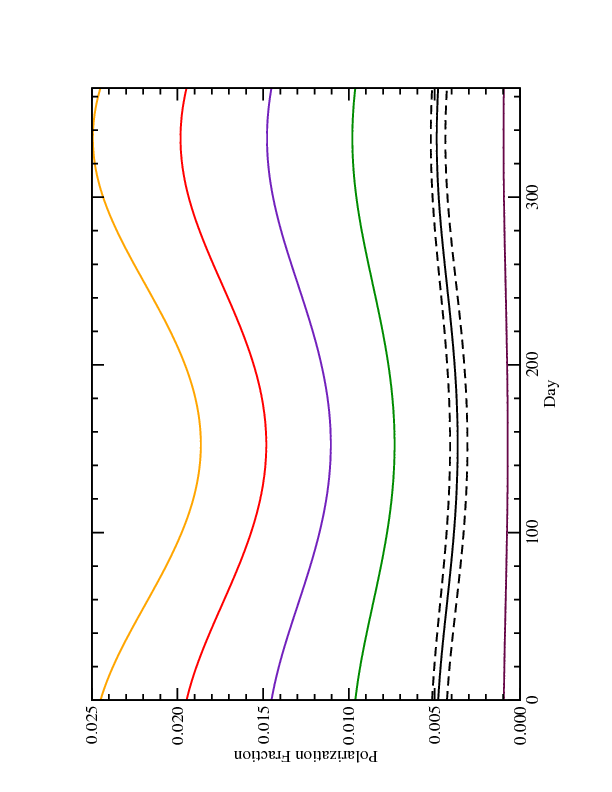}
\caption{\label{fig:pol} The polarization fraction $|{\cal P}|$ 
after Eq.~(\ref{fpol}) 
plotted as a function of day for various $v_{\rm stop}$
and $v_0$. 
Here $\bm{v}_E$ has been assumed parallel to 
$\bm{B}_0$; this yields a geometry-independent lower
bound to the polarization, as described in text. 
The solid lines have $v_0=220$ km/s and 
$v_{\rm stop}=1,5,10,15,20$, and $25$ km/s, respectively,
as one moves in the direction of increasing $|{\cal P}|$. 
The dashed curves have $v_{\rm stop}=5$ km/s and 
$v_0=200$ km/s and $v_0=240$ km/s, 
below and above, respectively, 
the solid line with the same $v_{\rm stop}$.
}
\end{figure}
We now proceed to estimate the polarization. 
We assume that the magnetic field is 
uniform in some direction $\hat{\bm{x}}$. 
If this is realized in a finite volume, 
with no magnetic field external to it, 
then a longitudinal 
Stern-Gerlach effect exists at each field boundary
--- i.e., the ``wrong'' spin state suffers a repulsive force
at each surface. In what follows we assume a slab geometry 
and estimate the polarization resulting from 
crossing a single 
interface; we assume 
the magnetic field is
perpendicular to the interface. 
In this case the polarization condition is on 
$\bm{v}_M\cdot\hat{\bm{B}}_0 \le v_{\rm stop}$, where $v_{\rm stop}$ is
fixed by Eq.~(\ref{vstop}). 
Thus the fraction of particles which enter 
the magnetic field region with 100\% polarization is 
\begin{equation}
f_{\rm pol} = \frac{1}{2} \int d^3v_M \,
f(\bm{v}_M, \bm{v}_E) \Theta ({v}_{\rm stop}
- |\bm{v}_M\cdot \hat{\bm{B}}_0|) \,, 
\label{filter}
\end{equation}
so that its polarization 
in the magnetic field region 
is ${\cal P} = ({\rm sgn}\, \mu) f_{\rm pol}/(1 - f_{\rm pol})$, and its 
number density is 
$n_M  = \rho(1 - f_{\rm pol})/M$. The evaluation of
Eq.~(\ref{filter}) yields two pieces, which are
distinguished by whether $|\bm{v}_M|$ is larger or
smaller than $v_{\rm stop}$. The term with 
$|\bm{v}_M|\le v_{\rm stop}$ 
can be evaluated irrespective 
of the orientation of the magnetic field region, 
whereas the term with $|\bm{v}_M|\ge v_{\rm stop}$ depends
on the value of $\bm{v}_E \cdot \bm{B}_0$. 
With Eq.~(\ref{maxwell}), we see that $f$ is maximized
for $\bm{v}_M$ in the neighborhood of $-\bm{v}_E$, so that
if $v_{\rm stop} \ll v_E$, the term with 
$|\bm{v}_M|\ge v_{\rm stop}$ dominates $f_{\rm pol}$, and 
the relative orientation of $\bm{v}_E$ and $\bm{B}_0$ becomes 
important. If, moreover, we make $\bm{v}_E\perp \bm{B}_0$, then 
we can have both 
$\bm{v}_M = - \bm{v}_E + \bm{\delta}$ and 
$\bm{v}_M \cdot \hat{\bm{B}}_0 =  \bm{\delta}\cdot \hat{\bm{B}}_0$ 
with $\delta$ small --- we expect $f_{\rm pol}$ to be 
maximized for this geometry. Thus to set a limit on $\mu$
irrespective of the orientation of $\bm{v}_E$ and $\bm{B}_0$, 
we choose $\bm{v}_E\parallel \hat{\bm{B}}_0$, as this geometry
gives the smallest value of $f_{\rm pol}$ for fixed $\mu$ and $M$. 
Ultimately the assessment of the limit on $\mu$ in a real
experiment will depend on the geometry and orientation of
the magnetic field, but our procedure should bound 
from above 
the limit on $\mu$ to be found from a given sensitivity
to the Faraday rotation angle per unit length. 
In this, we implicitly assume that dark matter is described
by a single constituent. 
In this special case we find 
\begin{equation}
f_{\rm pol}^\parallel = \frac{1}{4} 
\left(
\hbox{erf}\left(\frac{v_{\rm stop} - v_E}{v_0}\right) + 
\hbox{erf}\left(\frac{v_{\rm stop} + v_E}{v_0}\right) 
\right) 
\,.
\label{fpol}
\end{equation}
Since 
$\hbox{erf}(z) \to 1$ as $z\to \infty$, $f_{\rm pol}\to 1/2$ 
as $v_{\rm stop}\to \infty$, as required. 
The form of Eq.~(\ref{fpol}) emerges from a 
partial cancellation of the contributions from the
$v_M \le v_{\rm stop}$ and $v_M \ge v_{\rm stop}$ regimes, 
so that we pause to consider whether 
the inclusion of an escape velocity in
this particular case could modify our results. 
In this event, the normalization of 
Eq.(\ref{filter}) changes slightly, but negligibly~\cite{Lewin:1995rx},
and the integrand accrues a factor 
of $\Theta(v_{\rm esc} - |\bm{v}_M + \bm{v}_E|)$. 
For $v_{\rm stop} \ll v_E$, 
this additional factor does not restrict the
region of integration unless $v_M \gtrsim 600$ km/s; 
finally, we conclude that the continued neglect of $v_{\rm esc}$ is justified. 
Solving Eq.~(\ref{vstop}), we find 
$v_{\rm stop} \approx 4.51 \hbox{km/s}\,\sqrt{\kappa B_0[T]} (m_e/M)$,
where $B_0[T]$ is in tesla and $m_e$ is the electron
mass. Employing a 7 T magnet, as 
used in the UCNA experiment~\cite{ucna}, though 
magnets of up to 20 T are commerically available~\cite{oxinst}, 
we find
for $M=m_e$ and $\kappa=1$, with $\mu = \kappa \mu_M$, that
$v_{\rm stop}\approx 6$ km/s. 
The polarization fraction $|{\cal P}|$ 
as a function of day, using 
$f_{\rm pol}$ from Eq.~(\ref{fpol}), 
is shown for various  $v_{\rm stop}$ and $v_0$ 
in Fig.~\ref{fig:pol}. 
Day-by-day variations in the polarization exist 
since $f_{\rm pol}$ grows larger as $|\bm{v}_E|$ decreases. 
The time variation is more marked for the 
$\bm{v}_E\parallel \bm{B}_0$ geometry we have chosen, 
as the precise value of $\bm{v}_E$ impacts 
the range of velocities which enter the magnetic field region, 
and, hence, the polarization. 
For definitenss, we choose day 335 to set limits on $\mu$. 
The experiment we consider 
can be sensitive to both annual and daily signal 
variations, in principle. This follows from the duration of the 
photon--dark-matter interrogation time, which, in turn, is set by 
the size 
and finesse of the cavity in which the experiment is realized. 
We note that the total travel length $l$ of the laser light
in the 
PVLAS experiment is $l=4.4\cdot 10^6\,\hbox{cm}$~\cite{Zavattini:2005tm}, 
which corresponds to an interrogation time of 
${\cal O} (0.1\,\hbox{msec})$. 
This admits the possibility of using pulsed magnetic fields, 
which can be much stronger --- as much as 89 T~\cite{NHFML}. 
Irrespective of this, studies with the magnetic field on 
and then off are important to establishing a non-zero signal, 
if it is present. 
\begin{figure}
\includegraphics[scale=0.5,angle=-90]{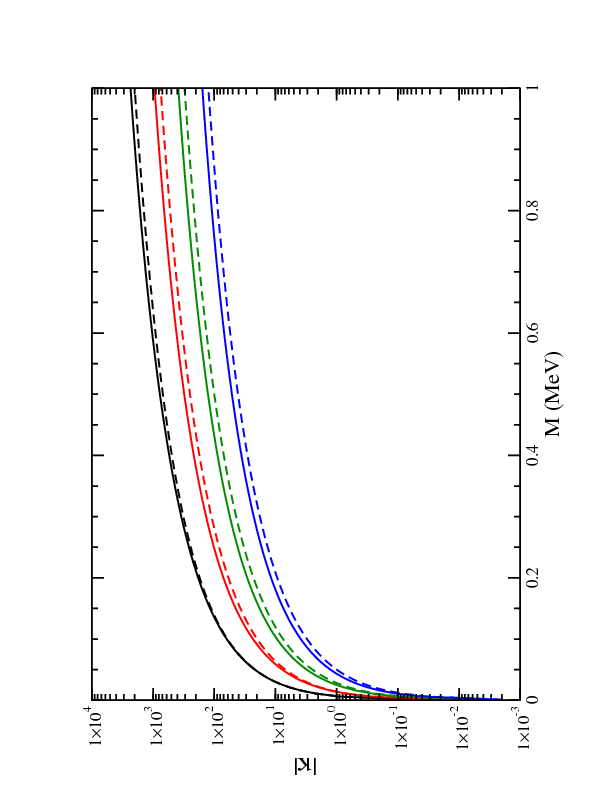}
\caption{\label{fig:limit} The limit on $|\kappa|$ at
95\% confidence interval, 
where $\mu = \kappa \mu_M$ and $\mu_M=e\hbar/2M$, 
as a function of $M$ up to 1 MeV 
for various limits on $\phi_0/l$ 
and for different values of the magnetic field $B_0$. 
The solid lines correspond to $B_0=7$ T, whereas
the dashed lines correspond to $B_0=20$ T.
The limit on $\phi_0/l$ in each case is $10^{-12}, 10^{-13}, 10^{-14}$, and
$10^{-15}$ rad/m, respectively, at 95\% confidence interval, 
as one sweeps from the top to the bottom of the figure. 
}
\end{figure}
\begin{figure}
\includegraphics[scale=0.5,angle=-90]{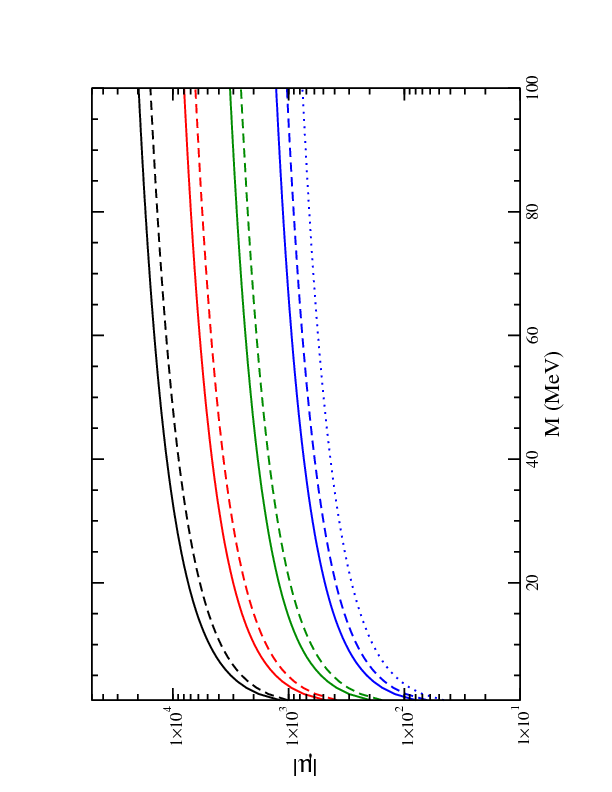}
\caption{\label{fig:biglimit} The limit on $|\mu|$ in units of $\mu_B$ at
95\% confidence interval, 
as a function of $M$ for masses ranging from 1 to 100 MeV 
for various limits on $\phi_0/l$ 
and for different values of the magnetic field $B_0$. 
We employ the notation of Fig.~\ref{fig:limit} throughout, and
note that dotted line corresponds to $B_0=89$ T and a limit
on $\phi_0/l$ of $10^{-15}$ rad/m at 95\% confidence interval. 
}
\end{figure}

\begin{figure}
\includegraphics[scale=0.5,angle=-90]{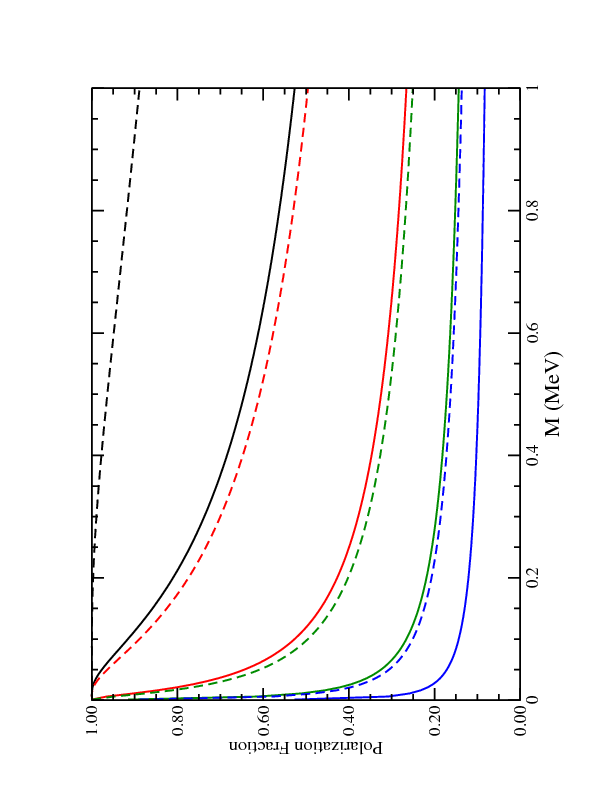}
\caption{\label{fig:fpollimit} The assessed polarization fraction 
$|{\cal P}|$ 
associated with the limiting value of $|\kappa|$ for fixed $M$ shown in 
Fig.~\ref{fig:limit}. 
We assume, as in Fig.~\ref{fig:pol}, that $\bm{v}_E\parallel \bm{B}_0$, to 
yield a geometry-independent lower bound to the polarization. 
The conditions which specify the
various curves are as given in Fig.~\ref{fig:limit}. 
}
\end{figure}

We now consider how a limit on the magnetic moment follows
from our determined polarization and a given sensitivity
to the Faraday rotation angle per unit length. For a spin-$1/2$ 
candidate, the rotation angle of Eq.~(\ref{phi0}) can be written as
\begin{equation}
\phi_0 = {\cal P}\kappa^2 \left( \frac{m_e}{M} \right)^3 \mu_B^2  \frac{n_e l}{\hbar c} 
\approx 
(6.84 \cdot 10^{-23} \hbox{cm}^2 )n_e[\hbox{cm}^{-3}]l[\hbox{cm}] {\cal P}\kappa^2 \left( \frac{m_e}{M} \right)^3  \,,
\label{phi0num}
\end{equation} 
with $n_e \equiv \rho(1- f_{\rm pol})/m_ec^2$
and ${\cal P}$ from $f_{\rm pol}$ as per Eq.~(\ref{fpol}). 
We note, too,
that if dark matter were a mixture of particle and antiparticles
of the same mass that the rotation angle could cancel, at least
in part. We can rewrite 
Eq.~(\ref{phi0num}) in terms of a limit on $|\mu|$ 
by replacing $\kappa m_e/M$ with $\mu [\mu_B]$. 
The most stringent limits on 
$|\kappa|$ emerge for $M \ll m_e$. 
For fixed $M$, its
numerical value rests on the ability 
to determine $\phi_0/l$. In the PVLAS experiment~\cite{Zavattini:2005tm}, 
the error in 
$\phi_0/l$ is determined to be 
$0.5 \cdot 10^{-12}$ rad/m. 
Since $l=4.4\cdot 10^4\,\hbox{m}$, $\phi_0$ itself
is determined to $2.2\cdot 10^{-8}$ rad. Assuming
then that $\phi_0/l$ can be determined to 
$1 \cdot 10^{-12}$ rad/m at 95\% confidence interval, 
we find the limit on $|\kappa|$, or $|\mu|$, as a function of $M$. 
In Fig.~\ref{fig:limit} we show the limit on $|\kappa|$ for 
candidate masses up to 1 MeV, whereas in Fig.~\ref{fig:biglimit}
we show the limit on $|\mu|$ for masses from 1 to 100 MeV. 
The limits depend 
on the dark-matter polarization ${\cal P}$ as well. 
To illustrate the relative importance
of ${\cal P}$ and the determination of $\phi_0/l$ 
to the limit on $|\kappa|$, we show not only how
the limits change if $B_0$ is increased from 
$7$ to $20$ T but also, in Fig.~\ref{fig:fpollimit}, 
the value of $|{\cal P}|$ associated with each limiting value 
of $|\kappa|$ in Fig.~\ref{fig:limit}. 
The increase in $B_0$ makes little
difference at the lightest mass scales we consider, 
simply because the polarization at these scales is already 
near unity. Indeed, increasing the value of $B_0$ 
simply increases the largest value of $M$ for which
we can reasonably constrain the value of $|\kappa|$. 
Our assessment of the polarization as per Eq.~(\ref{fpol})
is that of a lower bound, yet the polarizations we
find are large enough that our upper bounds on $|\kappa|$
are no more than a factor of a few larger than what
we would find after a realistic simulation of the 
geometry and orientation of the magnetic field region. 
It thus emerges that the limits to be set 
depend overwhelmingly on the ability to determine
$\phi_0/l$. Let us consider then the determination
of this quantity and its consequences carefully. 

In Figs.~(\ref{fig:limit}) and (\ref{fig:biglimit}) 
we show how the limits improve as the determination of
$\phi_0/l$ improves by orders of magnitude. This can be realized
by either increasing 
$l$ or by bettering the measurement of the rotation angle. 
We note, in particular, that better determinations
of $\phi_0$ are possible~\cite{budker}, and indeed that precision 
polarimetry at the shot-noise limit 
has been demonstrated~\cite{budkim,Budker:2002zz}. 
In this limit the error in $\phi_0$ is set, crudely, by
the number of photons counted, 
$\delta \phi_0 \simeq 1/(2\sqrt{I_o T})$, 
where $I_0$ is the number of photons per second and $T$ is the
measurement time. Assuming a $1$ W laser in the optical regime, so that
$E_\gamma \approx 1$ eV, we have 
$\delta \phi_0 \simeq 2 \cdot 10^{-10}\,
\hbox{rad-s}/\sqrt{T[s]}$~\cite{budker,Budker:2002zz}. 
Moreover, it has been demonstated that 
the use of squeezed light makes it possible to evade 
this quantum limit and realize yet more precise 
polarimetry~\cite{squeeze,Budker:2002zz}.
All this suggests that the significant gains in the determination
of $\phi_0/l$ we consider and more are indeed possible. 
Interestingly, as the determination of 
$\phi_0/l$ becomes more precise, increasing the value of 
$B_0$ becomes more important  to an improved limit on $|\kappa|$. 
We note, e.g., that if $M=100$ MeV and $B =89$ T 
that the polarization
fraction associated with the value of $|\mu|$  which 
follows from a limit of $\phi_0/l$ of $10^{-15}$ rad/m at
95\% confidence level is $|{\cal P}| \approx 0.09$. 

We have determined 
the direct limits on $|\mu|$, or $|\kappa|$, 
which would follow from a non-observation of Faraday rotation
at a given sensitivity. 
As they stand 
the limits constrain the possibility that the
dark-matter particle is a composite built from 
electromagnetically charged constituents. 
Significantly more severe limits would be needed to challenge
the possibility that a dark-magnetic magnetic
moment exists by dint of quantum-loop effects. 
Thus we wish to consider the prospect of 
radical improvement in our limits. 
Improvements can come from bettering the determination 
of $\phi_0/l$, increasing $B_0$, or, finally, noting 
Eq.~(\ref{phi0num}), from increasing the number density $n_e$. 
The ability to limit the
value of $\phi_0/l$ beyond that established
by the PVLAS experiment~\cite{Zavattini:2005tm}
has already been demonstrated~\cite{budkim,Budker:2002zz}, 
and the ultimate limit to the determination of $\phi_0/l$ has not yet
been established~\cite{squeeze}. Moreover, 
as discussed in Sec.~\ref{faraday}, mounting the Faraday 
rotation experiment in matter can yield 
magnetic fields which are larger by orders of magnitude, and
can possibly realize larger values of the 
dark-matter polarization ${\cal P}$ at fixed 
$\mu$ and $M$. Figures \ref{fig:limit} and \ref{fig:fpollimit} 
give a sense of the outcome 
as a result of such improvements. Let us now turn to the last
possibility. 

For the usually assumed Galactic halo density, 
$n_e$ is only some $600\,\hbox{cm}^{-3}$, so that the number
densities associated with warm dark matter are still very
low indeed. We wish to consider whether the value of $n_e$ can yield
to experimental manipulation. In particular, we would like 
to increase the dark-matter number density in a particular
region of momentum space. At first glance this would seem 
impossible~\cite{golub}, because 
Liouville's theorem demands that the density of the phase-space 
fluid for a system governed by a Hamiltonian is a constant of motion. 
Yet the Maxwell-Boltzmann velocity distribution we assume, cum 
gravitational potential, is a solution of Liouville's equation, 
so that a modest increase of number density can be realized
by reducing the gravitational potential. 
Significant gains,
however, are also possible, if inelastic processes operate, 
as in this case Liouville's theorem no longer limits the density. 
Such considerations are crucial to the construction of 
superthermal, ultra-cold-neutron sources~\cite{golub}; however, 
adapting this technology to our current context
does not appear to be practical.

\section{Summary}
\label{summ}

A Faraday effect also exists for light transiting a dark medium of 
electrically neutral particles with non-zero magnetic moments 
in an external magnetic field~\cite{svg}.  
We have used this notion to describe a 
Faraday rotation experiment which can lead to the direct detection of
dark matter were it to possess a magnetic moment $\mu$. Alternatively, 
a null result can be used to limit the magnetic moment of a 
dark matter particle. The set-up involves an evacuated optical
cavity to which an external magnetic field has been applied
and through which the light makes multiple passes in the direction 
of the magnetic field. We assume the dark matter wind in the Earth rest
frame sweeps through the apparatus. 
The size of the Faraday rotation angle is proportional to 
the magnetization of the dark matter, so that it is important
to give the magnetic moments some net alignment. 
For particles with sufficiently low kinetic energy, the 
passage into the magnetic field region itself acts as a longitudinal
Stern-Gerlach device; the highest energy state in the magnetic
field can be barred from entering the apparatus, engendering
a net polarization. We note that such a technique has been 
used to polarize ultra-cold neutrons with near 100\% 
efficiency~\cite{Tipton:2000qi}. Employing the usual 
assumptions~\cite{golwala} concerning the mass density and velocity 
distribution of galactic dark matter employed in the
analysis of existing direct detection experiments, 
we have estimated, for fixed astronomical input, 
the polarization of the dark matter in the
magnetic field region 
as a function of its mass and magnetic moment, as well as of the applied
magnetic field. Given a local dark-matter mass density, 
our limits on $|\mu|$ then follow from the strength of the 
applied magnetic field $B_0$ and from the ability to measure
$\phi_0/l$, the Faraday rotation angle accrued per unit length.

We have studied a range of candidate
masses compatible with dark matter as a warm thermal relic, namely,
from $1$ keV to $100$ MeV. This window in candidate
masses is not accessible via other direct detection techniques
and thus is unique to our study. We find the strongest limits
on $|\mu|$ emerge at the lightest mass scales we consider. 
In setting our limits we have employed the specifics
of existing, related experiments as far as 
possible~\cite{Zavattini:2005tm,Tipton:2000qi}, though 
we note that the possibility of more precise polarimetry
has already been demonstrated~\cite{budkim,Budker:2002zz}. 
The sensitivity of the limits we obtain are such that
they constrain the possibility that dark matter is a composite
with a magnetic moment, akin to a stable neutron without its
strong interactions. Indeed this analogy has proven useful 
in adapting technology used to manipulate neutrons to our
current case. 
The technical limits of the polarimetry
measurements have not yet been established~\cite{squeeze,Budker:2002zz},
and mounting the experiment in matter with concomitant gains in
the applied magnetic field may prove feasible, 
so that we can ultimately expect better limits on $|\mu|$
than those found in this paper.

\section{{Acknowledgments}} 
I am grateful to Geoff Greene for helpful discussions and 
particularly for 
suggesting that dark matter candidates 
with a magnetic moment could be polarized with 
a longitudinal Stern-Gerlach device. I thank 
Dima Budker for helpful comments and correspondence 
and for bringing Ref.~\cite{SLR} to my attention. 
I also thank Jeff Nico and Tom Gentile 
for many tolerant discussions and much encouragement, as well as 
Bob Golub and Mike Romalis for helpful remarks. 
I would like to thank the SLAC theory group, the Institute for
Nuclear Theory, the Kavli Institute for Theoretical Physics, 
as well as the Center for Particle Astrophysics and Theoretical Physics
at Fermilab, for gracious  
hospitality and to acknowledge partial support from the 
U.S. Department of Energy under 
contract DE--FG02--96ER40989, 
and from the National Science Foundation under
Grant No. NSF PHY05-51164. 

\end{document}